# Drawing induced texture and the evolution of superconductive properties with heat treatment time in powder-in-tube in-situ processed MgB$_2$ strands

M.A. Susner[1], T.W. Daniels[1,2], M.D. Sumption[1], M.A. Rindfleisch[3], C.J. Thong[3], and E.W. Collings[1]

1. Center for Superconducting and Magnetic Materials, Department of Materials Science & Engineering, The Ohio State University, Columbus, OH 43210 USA.
2. Swanson School of Engineering, University of Pittsburgh, Pittsburgh, PA 15260 USA.
3. Hyper Tech Research, Inc. 539 Industrial Mile Rd. Columbus, OH 43228, USA.

**Abstract**

Monocore powder-in-tube MgB$_2$ precursor strands were cold-drawn and heat-treated (HT) at 600$^o$C and 700$^o$C for times of up to 71 hours and structure-property relationships examined. Drawing-induced elongation of the Mg particles led, after HT, to a textured macrostructure consisting of elongated fine polycrystalline MgB$_2$ fibers (or veins) separated by elongated pores. The superconducting transition temperature ($T_c$), critical current density ($J_c$) and bulk pinning force density ($F_p$) were correlated with the macrostructure and grain size. Grain size increased with HT time at both 600$^o$C and 700$^o$C. Critical current density and hence $F_p$ decreased monotonically but not linearly with grain size. Overall, it was observed that at 700$^o$C, the MgB$_2$ reaction was more or less complete after as little as 30 min.; at 600$^o$C, full reaction completion did not occur until 72 hrs into the HT. Transport, $J_{ct}(B)$ was measured in a perpendicular applied field, and the magnetic critical current densities, $J_{cm}^{\perp}(B)$ and $J_{cm}^{\varphi}(B)$, were measured in perpendicular and parallel (axial) applied fields, respectively. Particularly noticeable was the premature dropoff of $J_{cm}^{\perp}(B)$ at fields well below the irreversibility field of $J_{ct}(B)$. This effect is attributed to the fibrous macrostructure and its accompanying anisotropic connectivity. Magnetic measurements with the field directed along the strand axis yielded a critical density, $J_{cm}^{\perp}(B)$, for current flowing transversely to the strand axis that was less than and dropped off more rapidly than $J_{ct}(B)$. In the conventional magnetic measurement, the loop currents that support the magnetization are restricted by the lower of $J_{ct}(B)$ and $J_{cm}^{\phi}(B)$. In the present case the latter, leading to the premature dropoff of the measured $J_{cm}(B)$ compared to $J_{ct}(B)$ with increasing field. This result is supported by Kramer plots of the $J_{cm}^{\phi}(B)$ and $J_{ct}(B)$ data which lead to an irreversibility field for transverse current that is very much less than the usual transport-measured longitudinal one, $B_{irr,t}$.

**Introduction**

Many studies have been devoted to improving the superconducting properties of $MgB_2$ since its discovery in 2001 [1]. Superconducting powder-in-tube (PIT) strands have been made by various groups using the *in situ* method [2-10] (in which a mixture of Mg and B powders enclosed in a non-reactive tube is drawn to wire and then reacted to form $MgB_2$) and the *ex situ* method [6],[11-12] (in which the starting powders are pre-reacted to form $MgB_2$ which, after pulverization, is also enclosed in a tube and drawn to final strand). One advantage of the *in situ* process is that the Mg+B reaction, being facilitated by the relatively high vapour pressure of Mg, can in principle begin below the melting point of Mg (650$^o$C) [13]. In contrast, the inability of $MgB_2$ to sinter even under extreme conditions is well known [14] and requires *ex situ* processed wires to undergo final heat treatments of 800-1000$^o$C [11-12].

Much research into improving the superconducting properties of $MgB_2$ strands has focused on: (i) increasing the critical current density ($J_c$) through increasing the bulk pinning force density ($F_p = J_c$ x $B$) [15-20]; (ii) increasing the critical fields ($B_{irr}$ and $B_{c2}$) and hence the high-field $J_c$ through the introduction of dopants that contribute to charge-carrier scattering. These can be accomplished through: (i) site-substitution into the Mg sublattice by Zr [21, 22], Na [23], or similar-sized cations [24-27], (ii) site-substitution into the B sublattice by C [28-32] (with regard to which we note that it has been difficult to distinguish t-he influence of substitution from effect of possible C-induced lattice strain [33][34]), and (iii) increasing the connectivity between $MgB_2$ grains (or grain clusters) to increase the efficiency of supercurrent flow [35-38].

Increasing the electrical connectivity of the $MgB_2$ is the single most important route towards high $J_c$. Using the transport properties of $MgB_2$ single crystals as reference it has been shown that bulk pellets and PIT strands prepared under zero- to moderate externally applied pressures (the usual conditions) yield electrical connectivities of only 10-20% [27][39]. Moreover, $J_c$ has been shown to scale with the normal-state electrical connectivity, with more highly connected samples exhibiting higher critical current densities [40]. It has been suggested that poor connectivity results from the presence of intragranular insulating oxide layers; these in turn stem from an MgO coating on the Mg particles [41] or a $B_2O_3$ coating on the B particles [42]. These oxide layers are also thought to inhibit the sintering of $MgB_2$ particles during *ex situ* route processing by restricting intergrain atomic diffusion.

Connectivity manifests itself at two levels: (i) between the individual $MgB_2$ grains, (ii) between aggregates (fibers, see below) of even well connected polycrystalline grains. With regard to *in situ* processed PIT $MgB_2$ strands, it has been noted [4][43] that the drawing of the mixed powders leads eventually to a fibrous macrostructure. The term "macrostructure" is herein used to indicate that the fiber dimensions (typically 27 µm x 2.5 µm) are very much larger than the typical grain size ($\approx$ 50 nm). It has been shown qualitatively [4] that this fibrous microstructure begets an anisotropic connectivity. Numerous authors have noted a difference between the transport measured $J_{ct}$ and the magnetically measured $J_{cm}$s [4][44-45]. Some authors have claimed these differences to be due to defects [44] or current loops of different length scales caused by porosity [45]. This report offers an alternative model based on macrostructurally induced $J_c$ anisotropy. Magnetization measurements with the applied field directed along the strand axis yield an azimuthal critical current density $J_{cm}^{\phi}$ *(B)* that can be identified with a transverse (across-the-strand) critical current density $J_2$. It turns out that over the entire field range $J_2$ is less than $J_{ct}$ as are the corresponding irreversibility fields. In the conventional magnetic critical current density measurement, the loop currents that support the magnetization are restricted by the lower of $J_{ct}(B)$ and $J_2(B)$; in the present case this is $J_2(B)$. Application of an anisotropic critical current density model previously used for low temperature superconductors [46] is appropriate and explains the noted premature dropoff of $J_{cm}^{\perp}(B)$ with increasing field relative to $J_{ct}(B)$. For clarity, Table I lists and defines all critical current densities referred to in this work.

However, understanding the nature of the anisotropic $J_c$ and how it relates to the macrostructure and chemistry of the drawn $MgB_2$ wires is essential for establishing a predictable structure-property relationship. To this end, we have performed detailed analyses of the evolution of $T_c$, $J_c$, macrostructure and microstructure during HT for times of 0.5–1 hrs at 600°C and 700°C, temperatures that were chosen to bracket the 650°C selected by others (e.g. [47]) and which are also 50°C below and 50°C above the melting point of Mg. The following results show that the $MgB_2$ formation reaction proceeds to completion much more rapidly as the HT temperature is raised from 600°C to 700°C assisted by a 7-fold increase in Mg's vapour pressure (0.68 torr to 4.84 torr, respectively). Additionally, we have investigated the effect of drawing on the powder morphology in the *in situ* wires at various stages in the manufacturing process to explain the origin and development of the fibrous macrostructure and its influence on the anisotropic connectivity.

**Experimental Procedure**

Monofilamentary powder-in-tube (PIT) strands with a Nb chemical barrier and an outer sheath of Monel 400® (henceforth "monel", a copper-nickel alloy) were manufactured by HyperTech Research Inc. (HTR)  The final strand had a cross-section which was (by area) 50% monel, 29% Nb barrier, and had a remaining 21% allocated to the mixed-powder core. The basic powder ingredients were commercial Mg powder (99%, < 44 μm particle size) and "amorphous grade" B from Alpha Aesar. Listed as 99% pure, and made from the thermal decomposition of diborane gas, this powder consists of submicrometer-size agglomerations of 30 -100 nm size particles. Also included in the starting powder as a C-contributing dopant was small fraction of malic acid. (Alfa Aesar, 99% purity) which was mixed with the B powder in the ratio of 98.3 mol%B to 1.7 mol% malic acid. After this, 60 mL of toluene was added to create a slurry which was subsequently high energy ball milled for 12 minutes. The slurry was then transferred to a pyrex jar and vacuum-dried at 150$^{o}$C during which the malic acid melted and decomposed and, along with the toluene, contributed C to the powder mixture  [31] . An appropriate amount of Mg powder was then was added to restore the Mg:B ratio to 1:2. The powders were planetary milled at 100 rpm for 10 minutes followed by additional planetary milling steps performed at 500 rpm for 12 mins.   Bimetallic Nb/monel tubes were filled with this powder using HTR's previously described "CTFF" process [48-50] in preparation for wire drawing to a diameter of 0.83 mm followed by *in-situ* heat treatment (HT).

Segments of strand about 20 cm long, with ends crimped to prevent Mg loss, were heated in Ar at 6$^{o}$C/min  to  600$^{o}$C and 700$^{o}$C where they remained for times of 0.5, 1, 2, 4, 6, 8, and 71 hrs before being quenched in ice water. Transport critical current density $J_{ct}$ was measured on 3 cm long pieces cut from the centers of the heat treated (HT) strand segments. Measurements were performed at temperatures 4.2 K, in magnetic fields of up to 14 T at currents of up to 220 A. The gauge length was 5 mm and the $J_c$ criterion was 1 μV/cm.

Both $T_c$ and magnetic critical current densities in transverse and longitudinal fields of up to 14 T were measured on ~5 mm segments of strand using the VSM option on a Quantum Design Model 6000 PPMS after the monel outer sheath had been etched off using a 20% aqueous solution of $HNO_3$.

Ultra-high-resolution micrographs were obtained using a Philips Sirion field-emission gun (FEG) scanning electron microscope (SEM) with a through-the-lens detector (TLD). For the longitudinal cross-section micrographs, samples were cold-mounted in conductive epoxy and any bubbles were removed by pumping. These 1.5-inch diameter mounts were then sequentially polished using 320, 400, 600, 800, 100, and 1200 grit SiC polishing paper followed by 6 μm and 1 μm diamond polishing. For grain size measurement fracture SEM samples were prepared after etching off the monel outer sheath with the same procedure as above and fracturing the resulting $Nb/MgB_2$ strand. These samples were held in place using an Al clamp. Grain size was measured using the line-intercept technique.

**Results and Discussions**

*Macrostructural Properties: The Unreacted Strand*

It is well known that Mg has limited ductility, with an elongation-to-fracture of ~7% under static strain. However, a Mg rod embedded in fine B powder (either amorphous or microcrystalline) can undergo ductile elongation when drawn through a series of dies. This phenomenon is well-known, having been performed by both Giunchi [51,52] and Kumakura [53] who described packing an axial rod of Mg into an experimental billet with B powder and drawing it to fine wire sizes. Likewise imbedded Mg particles also undergo elongation during drawing though dies as reported previously by Uchiyama [43] and shown conceptually in Figure 1.

The elongation of the Mg particles during the drawing of an actual strand is shown in Figure 2. Depicted are longitudinal sections of a strand at successive stages of wire reduction as the outside diameter (OD) decreases from 4.08 mm to 0.83 mm. In these unreacted strands the aspect ratio of the Mg particles increased from 1 to 12 as the strand was drawn to smaller and smaller diameters, Figure 3.

*Macrostructural Properties: The Reacted Strand*

During HT a "shrinking-core reaction" Mg+B → $MgB_2$ takes place during which Mg liquid or vapour reactively diffuses into the B particles, forming $MgB_7$, $MgB_4$, and finally $MgB_2$ [53]. This is confirmed on the macroscopic scale by observations following the formation of $MgB_2$ in the "internal magnesium diffusion" (IMD) process [51-53]; wherein a void replicates the space previously occupied by Mg. Powder-in-tube *in situ* $MgB_2$ formation is analogous; its noted porosity is a direct result of the voids left behind after the prior Mg has reacted.

As shown in Figure 2 the pre-reacted strand consist of aspected stringers of Mg and correspondingly elongated regions of B powder. During HT the Mg leaves behind long, aspected voids as it reacts with the B powder to form a fibrous $MgB_2$ macrostructure. These $MgB_2$ fibers, or veins, which are typically ~20 µm long and ~2.5 µm across (the macrostructure) consist of an assembly of randomly oriented, relatively well-connected fine polycrystalline $MgB_2$ grains typically 50 nm in size (the microstructure). The veins of polycrystalline $MgB_2$ are preferentially oriented longitudinally within the strand. The electrical connectivity between them differs in transverse and longitudinal directions. To be discussed below is the profound effect of this anisotropy on the superconducting properties of the $MgB_2$ strands.

Figure 4 shows the evolution of the macrostructure as a function of HT time at 600°C and 700°C. Increasing the HT temperature from 600°C to 700°C, which increases Mg's vapour pressure from 0.68 torr to 4.84 torr, has a pronounced effect on the Mg+B → $MgB_2$ formation kinetics. In Figure 4 the pre-reacted strand clearly shows the aspected Mg particles that later give rise to elongated voids. At 700°C, the macrostructure is mostly composed of $MgB_2$ with occasional areas of a higher boride phase, likely to be $MgB_4$ or $MgB_7$ (seen as the darker regions in the back-scatter micrographs). These regions reduce in area with the HT time, albeit slowly. At 600°C, these higher boride phases dominate the micrographs. However, as the HT time is increased these regions decrease in size until, at 71 hrs, they occupy the same fraction of the strand as they do after the 700°C HT.

*Transition Temperature by DC Susceptibility*

The superconducting transition temperature, $T_c$, was determined by zero-field-cooled (ZFC) DC susceptibility vs. temperature ($T$) measurements on strands that had received HT for 0.5, 1, 8, and 71 hr at 600°C and 700°C. In the applied field of 200 Oe which is less than $MgB_2$'s 4 K lower critical field of ~250 Oe [55] the DC susceptibility responds to a Meissner screening current flowing within a penetration depth of 140–150 nm [55][56] which is of the order of a few grain diameters. So the DC susceptibility measurement tends to probe properties at the level of the MgB$_2$ veins. Specifically, any observed distribution in $T_c$s would reflect intra-fiber inhomogeneities rather than whole-strand inhomogeneity, if present. The magnetic results are presented in Figure 5 in the formats normalized susceptibility ($\chi/\chi_0$) vs $T$ and its derivative, $\partial(\chi/\chi_0)/\partial T$, vs. $T$.

Figure 5 shows that with increasing HT time at 600°C, $T_c$ increases gradually while the distribution sharpens, in response to increasing homogeneity (as the reaction gradually proceeds to completion). Indeed at 600°C it takes 8 hrs for the $T_c$ distribution to match that of the sample HT for only a short time at 700°C. During HT at 700°C, both $T_c$ and its distribution reach their saturation values in about one half an hour. The peaks seen within 7.5~9.5 K represent the $T_c$s of the Nb chemical barriers. The onset $T_c$s of the MgB$_2$ components from Figure 5 are plotted as functions of HT time in Figure 6. Evidently the formation reaction seems already complete after about one half an hour of HT at 700°C. while at 600°C at least 71 hrs seem to be required.

*Critical Current Density in Response to HT Time and Temperature*

Transport and magnetic critical current densities ($J_{ct}$ and $J_{cm}^\perp$, respectively) were measured at 4.2 K as functions of magnetic field applied perpendicular to the wire axis. For later use the longitudinal-field magnetic critical current density $J_{cm}^\varphi$ was also measured; the transport results are presented in Figure 7. At 700°C, $J_{ct}$ achieved $10^4$ A/cm$^2$ at 11-11.2 T independently of HT time in the range 0.5-2 hrs. Unfortunately, what we deduced as Nb barrier failure after more than 2 hrs of HT prevented reliable $J_{ct}$ measurements from being made on samples HT for longer times. At 600°C, $J_{ct}$ rose to $10^4$ A/cm$^2$ at 10.7 T after HT for 4 hrs. Longer times caused a slight drop-off in $J_{ct}$, to $10^4$ A/cm$^2$ 9.75 T after 6 hrs and 8 hrs of HT.

In order to compare the effects of long HT times at 600°C and 700°C on critical current density, bearing in mind that the 700°C $J_{ct}$ was unavailable for HT times longer than 2hrs, it was necessary to frame the comparison in terms of the transverse-field magnetic $J_{cm}^{\perp}$. In spite of the well known difference between $J_{ct}$ and $J_{cm}$ in *in situ* PIT MgB$_2$ strands [4][44-45] it is nevertheless useful to use $J_{cm}^{\perp}$ as a basis for comparison when comparing the effects of long-time HTs at different temperatures when $J_{ct}$ data are not available. The 5 K, 4 T values of $J_{cm}^{\perp}$ as functions of HT time at 600°C and 700°C are presented in Figure 8.

Recognizing that the formation reaction at 700°C is already complete after about one half an hour of HT we attribute the monotonic decrease in $J_{cm}^{\perp}$ with HT time to grain growth which reduces specific grain-boundary surface area hence the effective density of flux pins. At 600°C, $J_{cm}^{\perp}$ steadily increases with HT time until at 71 hrs it is almost equals that achieved by a 700°C HT for the same period of time (10.2 and 9.4 $10^4$A/cm$^2$, respectively). This result conforms with the observation that both of the 71 hr HT samples have the same macrostructure, similar $T_c$s, and similar $T_c$ distributions. More importantly, as we shall see in the next section, both the 71h/600°C and 71h/700°C HTs produce similar grain sizes.

*Grain Size, Grain Growth and Flux Pinning*

Ultra-high-resolution SEM was employed on fracture samples in order to gauge the effect of heat treatment parameters on the grain size, Figure 9. The grain sizes, extracted from the SEMs using the line-intercept technique and are presented in Figure 10 as a function of HT time, $t$. During the 700°C HT the MgB$_2$ grain size, $d$, increases proportionately to $t^{1/2}$ (albeit at an extremely slow rate) in conformity with the standard treatment of grain growth according to which $d^2-d_0^2=Kt$ [57] in which $d_0$ is the initial grain size (38 nm in this case) and $K$ is the rate constant. However, after 4 hrs of HT no further growth takes place with the grain size reaching a plateau of 51.7 ± 6.4 nm. During the 600°C HT, the grain size remains constant up to about 4 hrs during which the Mg+B → MgB$_2$ approaches completion (see the magnetic susceptibility and $T_c$ behaviours shown in Figures 5 and 6). Once the reaction is more or less complete (although as evidenced from the d$M$/d$T$ graph in Figure 5 not necessarily homogeneous) the MgB$_2$ grain growth at 600°C follows the same $t^{1/2}$ dependence as before. Once more, after a relatively short period of limited grain growth (from $d$ = 36.8 ± 2.3 nm at 4 hrs to 50.0 ± 3.2 nm at 8 hrs), grain a prolonged HT of 71 hrs at 600°C and 700°C the grains reached sizes of 53.2 ± 3.9 nm and 53.2 ± 4.3 nm, respectively, and stop growing.

These similar final grain sizes suggest that grain growth is inhibited by a common mechanism the most likely candidate being grain-boundary contamination which eventually blocks the intergranular diffusion needed for growth. Numerous authors have conducted detailed studies of grain boundary contamination [58], some emphasizing its effect on electrical connectivity [59]. Boron powder when exposed to air (as usual in in-situ PIT processing) readily acquires a coating of B$_2$O$_3$. Early in the HT of the Mg+B mixture the B$_2$O$_3$ melts (M.P., 450°C), coats whatever particles are present, in particular the Mg particles with which it reacts to form MgO. It has been shown that the final MgB$_2$ grains end up partially coated with an insulating MgO film that by acting as a current barrier, reduces the effective grain-surface area, and restricts the electrical connectivity of the polycrystalline superconductor [59]. We postulate here that this same partial coating restricts, and eventually stops, the growth of the $MgB_2$ grains during HT.

We conclude this section on granularity by enquiring into the relationship between HT time, grain gowth, $J_{cm}$, and bulk pinning force density with particular reference to the 700°C HT (since at this temperature the strands were already fully reacted after 0.5 ht). Figure 11(a), which juxtaposes the $J_{cm}^{\perp}$ (5K, 4T) data of Figure 8 against the grain size data of Figure 10, shows the former monotonically decreasing and the latter increasing with HT time at 700°C. Figure 11(b) shows the 5-K, 4-T bulk pinning force density $F_p$ (equal to $J_{cm}^{\perp}$ x B) plotted against inverse grain size. It is generally accepted by now that $MgB_2$ is a normal-surface (grain-boundary) pinner whose $F_p$ follows the well known Kramer-Dew-Hughes relationship [60, 61]. In ideal grain-boundary pinning one would expect $F_p$ to be proportional to the grain surface area per unit volume and hence to *1/d*. Although the present sample deviates from this ideal geometry-based relationship, $F_p$ clearly increases with decreasing *d*, a result previously seen in $MgB_2$ wires, tapes, and films [62].

*Transport- and Magnetic Critical Current Densities*

Comparisons were made of the transport and magnetic $J_c$s in perpendicular applied fields ($J_{ct}$ and $J_{cm\perp}$, respectively) of strands HT for 2 hrs at 600°C. This HT was chosen since it enabled a direct overlap of $J_{ct}$ and $J_{cm}^{\perp}$; the strand HT at 700°C would have had a low field critical current greater than the 120 A limit of the power supply. For the magnetic measurements the samples were cut to two lengths corresponding to aspect ratios (length/diameter, *L/D = S*) of 13 and 8. Associated with Figure 12, are several important observations: (1) The transport $J_{ct}$ extends well past the equipment limit of 14 T; (2) The magnetic $J_{cm}^{\perp}$ dies off "prematurely" at a rate that depends on $S$, indeed $J_{cm}^{\perp}$ plunges towards zero in fields of 7 – 11 T; and (3) The $J_{cm}^{\perp}$s of $J_c$-isotropic cylinders with $S$-values of 13 and 8 differ from $J_{ct}$ by only about 2% and 4%, respectively (see Equation 1(a) below).

A relationship between the drop-off of $J_{cm}^{\perp}$, the $S$-dependence of that drop-off, and departure from $J_c$-isotropy (i.e. anisotropy) is discussed below with reference to the results of detailed SEM observations and critical state analysis. In a perpendicular applied field the current loop supporting the magnetization consists of longitudinal currents of density, say $J_3$, (which can be identified with $J_{ct}$) in series with transverse currents of density, say $J_2$. According to Sumption's calculations [46] the "perpendicular field" magnetic critical current density, $J_{cm\perp MOD}$, is sample-aspect-ratio dependent and given by:

$$J_{cm,MOD}^{\perp} = J_{c3}\left[1 - \frac{3\pi D}{32L}\left(\frac{J_{c3}}{J_{c2}}\right)\right] \qquad L_t < \frac{L}{2} \qquad (1a)$$

$$J_{cm,MOD}^{\perp} = \frac{3\pi L}{8D}J_{c2}\left[1 - \frac{4L}{3\pi D}\left(\frac{J_{c2}}{J_{c3}}\right)\right] \qquad L_t > \frac{L}{2} \qquad (1b)$$

in which $L_t$ represents a quantity which is analogous to the current transfer length of an anisotropic finite-slab model [46]. Associated with Figure 12 and Equation (1) are several important observations: (i) The transport $J_{ct}$ is $10^3$ A/cm$^2$ at 12 T while the magnetic $J_{cm}^{\perp}$s drop off "prematurely" at a rate that depends on $S$ and fall to $10^3$ A/cm$^2$ at 7.9 T ($S$ = 13) and 6.9 T ($S$ = 8). (ii) For long isotropic superconductors ($L \gg D$, $J_{c3} = J_{c2}$) Equation (1a) shows that $J_{cm\perp MOD} \approx J_{c3} \equiv J_{ct}$. Evidently the observed $J_{ct}$ - $J_{cm}^{\perp}$ bifurcation can be attributed to finite aspect ratio coupled with critical current anisotropy. The latter seems to be the driving mechanism since, according to Equation (1a), the $J_{cm\perp MOD}$s of isotropic cylinders with *S*-values of 13 and 8 differ from $J_{ct}$ by only about 2% and 4%, respectively, scarcely the thickness of the lines in Figure 12.

The projected deviation between $J_{cm\perp MOD}$ and $J_{ct}$ becomes even stronger once field dependencies are introduced, however the modification of Equation (1) needed to include the field dependencies of $J_3$ and $J_2$ is beyond the scope of this work. Figure 12 compares the measured values $J_{ct}$ and $J_{cm}^{\perp}$. Next, in order to compare $J_{ct}$ with the model $J_{cm\perp MOD}$s predictions of Equation (1) it is necessary to find $J_2$. This we do by applying the field along the strand axis and from the resulting magnetization determine the circumferential critical current density, $J_{cm}^{\phi} = 30\Delta M/D$ [63]. We then take the next step of identifying $J_{cm}^{\phi}$ with $J_2$ since both currents pass transversely across the strand and encounter the same microstructural features. After inserting $J_{cm}^{\phi} \equiv J_2$ into Equation (1) and identifying $J_3$ with $J_{ct}$ it is possible to calculate the model $J_{cm\perp MOD}$ and compare it with the measured $J_{cm}^{\perp}$ as in Figure 13. Also included in that figure are (i) the transport-measured $J_{ct} \equiv J_3$, (ii) the magnetically-measured $J_{cm}^{\phi} \equiv J_2$, and (iii) Kramer plots based on $J_{ct}(B)$ and $J_{cm}^{\phi}(B)$.

In Figure 13, although $J_{cm\perp MOD}(B)$ and $J_{cm\perp}(B)$ do not track exactly, they both tend towards a common irreversibility field ($B_{irr}$, $10^2$ A/cm$^2$ criterion) of about 10 T; the $B_{irr}$ associated with $J_{ct}(B)$ is projected to be about 16 T. Referring again to Figure 13 the differences between $J_{ct}(B)$ and $J_{cm}^{\phi}(B)$ (i.e $J_3$ and $J_2$) and the differences between the corresponding $B_{irr}$s are clearly responsible for the bifurcation of the measured $J_{ct}(B)$ and $J_{cm\perp}(B)$.

Evidently the drop-off of $J_{cm\perp}(B)$, the $S$-dependence of that drop-off, and $J_c$-anisotropy are interrelated. Several authors have reported on the influence of sample size, at fixed $S$ [64,65], and fixed and variable $S$ [66] on magnetic-$J_{cm}$. Attention was directed primarily on measurements with the field applied parallel to the sample's long axis and various explanations for the size dependencies were offered. In a broad extension of the size-effect studies Horvat [45] compared the field dependencies of $J_{ct}$ and $J_{cm}$ in PIT-processed $MgB_2$ strands, attributing observed differences to macrostructural properties (in present terminology) of the superconducting core – porosity and agglomerations of superconducting crystals. Our model also includes these basic elements but is bolstered on one hand by critical state analysis and on the other by detailed SEM observations.

*Macrostructural Anisotropy and its Influence on Supercurrent Flow*

The macrostructure of the reacted PIT strand core is characterized by an array of ~20 x ~2.5 µm polycrystalline $MgB_2$ fibers and their associated elongated pores aligned along the wire direction, Figure 4. The existence of a $J_3$ and a $J_2$ indicates that the fibers are both longitudinally and transversely interconnected. The presence of anisotropic critical current density was initially surprising, given that the $MgB_2$ is polycrystalline with fine randomly oriented grains, Figure 9. However, in light of the evolution of fibrous stringers of Mg during the wire drawing process, we propose the following explanation for the observed electrical anisotropy in $MgB_2$ strands.

The macrostructure of the as-drawn but unreacted strand is depicted in Figure 14a which shows the B as the dark phase and the Mg as the white phase. Evidently, as the strand is drawn down to its final diameter of 0.83 mm, the Mg elongates and fractures, forming isolated, aspected stringers (the white regions in Figure 14). It is well-known that Mg readily oxidizes in air. Accordingly an MgO film will surround every starting Mg particle. As the Mg elongates with drawing the brittle film will fracture leaving a partially coated surface. As described above, further surface oxidation will take place during HT.

The B powder will be forced to rearrange itself into veins lying continuously along the length of the strand, Figure 14. The Mg vapour that evolves during HT permeates the B and begins to convert it to veins of $MgB_2$. As a result of the density difference between B and $MgB_2$ [67] these veins will swell during this conversion (resulting in a 1.2-fold increase in volume assuming an initial B density of 65%) and will stop growing when either all the B is converted to $MgB_2$ or until neighboring veins of $MgB_2$ impinge upon each other. Just as in Giunchi's "reactive magnesium infiltration process" [51,52] for producing $MgB_2$ bulks and strands, each reacting Mg stringer leaves behind a matching vacancy. As a result these elongated pores that separate the veins of $MgB_2$ produce a relative weakening of the transverse connectivity. Impurity phases such as MgO, originating from the starting Mg particles and as a product of oxidation by $B_2O_3$ during the HT [58,59], also decorate the surface of the $MgB_2$ veins and again contribute to the anisotropic connectivity. This explanation is presented pictorially in Figure 14, for clarity in terms of the pre-reacted structure (cf. Figures 2 and 4): (i) Figure 14(a) shows the locations of the elongated B and Mg sites. (ii) In Figure 14(b) long dashed lines emphasize the continuous nature of the longitudinal B (future $MgB_2$) veins. (iii) In Figure 14(c) short dashed lines emphasize the transverse B connections showing them to be both discontinuous and circuitous, leading to poor transverse connectivity in the HT strand, and hence the large difference between the longitudinal and transverse critical current densities, $J_{ct}(B)$ and $J_2(B)$.

**Concluding Summary**

The structure-property relationships in a superconducting $MgB_2$ strand have been investigated. The superconducting properties $T_c$, magnetic- and transport critical current densities, $J_{cm\perp}(B)$ and $J_{ct}$, and bulk flux pinning force density, $F_p$, were measured and correlated with processing induced materials properties. The evolution of cold drawn microstructure from a mixture of Mg and B powders to a macrocomposite of elongated Mg stringers in a correspondingly elongated B matrix was observed. Next, during HT for times of 1-71 hrs at 600°C and 700°C, the conversion of the B component of the mixture to $MgB_2$ fibers separated by elongated pores (the prior Mg sites) was observed. The speed of the $Mg+B \rightarrow MgB_2$ reaction at 600°C and 700°C was studied by magnetic- and transport-property measurement. Magnetic measurements of $T_c$ and transport measurements of $J_{ct}$ showed the reaction to be barely complete after 71 hrs at 600°C but fully so after only about 0.5 hrs at 700°C. With regard to the detailed structure of the strand: (i) SEM performed before HT showed the aspect ratio of the Mg stringers increasing from 1 to 12 as the area reduction of the strand increased up to 90%; (ii) SEM performed "during" HT showed the evolution of macroscopic veins of $MgB_2$ separated by elongated pores; (iii) fracture-SEM depicted a fine equiaxed randomly oriented granular structure within the veins, and a grain size, $d$, that increased from 36 ~ 52 nm with HT time, $t$. Although the rate of grain growth was found to follow the usual $d \approx K\ t^{1/2}$ relationship, the rate constant $K$ which was the same for both HT temperatures (albeit after a delayed start at 600$^{\circ}$C) was relatively small. It is postulated that grain growth is inhibited by the presence of grain boundary contaminants such as MgO, the same contaminants that reduce inter-grain (i.e *within* the MgB$_2$ veins) connectivity. After correlating grain size with $J_{cm\perp}$ it was noted that although an increase with decreasing grain size was observed, the rate of increase was not directly proportional to the $d^{-1}$ generally expected for a dense packing of equiaxied grains, probably again due to the intervention of grain-boundary film. An important component of the research was a comparison between $J_{cm\perp}(B)$ and $J_{ct}(B)$ accompanied by an explanation for the "premature" drop-off of $J_{cm\perp}(B)$ relative to $J_{ct}(B)$. The transverse-field magnetization giving rise to $J_{cm\perp}(B)$ is supported by a strand-longitudinal critical current density $J_{ct}(B) \equiv J_3(B)$ in series with a transverse component, $J_2(B)$. As mentioned above, the core of the strand consists of MgB$_2$ fibers separated by elongated pores. The fibers are practically continuous along the length of the strand but are tortuously interconnected transversely to the wire drawing direction. These degraded interconnects, most likely contaminated with MgO, cause the $J_2(B)$ to be both weaker than more strongly field dependent than $J_{ct}(B)$ thereby driving a weaker and more strongly field dependent $J_{cm\perp}(B)$. As a corollary we note that in characterizing the critical current density of an *in-situ* PIT strand the magnetically determined value is not a reliable substitute for the transport-measured $J_{ct}$.

**Acknowledgements**

This work was supported by the United States Department of Energy, Office of High Energy Physics Grants DE-SC0001546 and DE-FG02-95ER40900, the United States National Institutes of Health Grant 1RC3EB011906-01, as well as a grant from the Ohio Department of Development.

**References**

[1]   J. Nagamatsu, N. Nakagawa, T. Muranaka, Y. Zenitana, and J. Akimitsu, *Nature* **410,** 63-64 (2001).

[2]   M. Bhatia, M.D. Sumption, M. Tomsic and E.W. Collings, *Physica C* **407** (3-4) 153-159 (2004).

[3]   M.D. Sumption and E.W. Collings, *Physica C* **401** 66-74 (2004).

[4]   Z. X. Shi, M.A. Susner, M. Majoros, M.D. Sumption, X. Peng, M.A. Rindfleisch, M.J. Tomsic, and E.W. Collings, *Supercond. Sci. Technol.* **23**(4), 045018/1-10 (2010).

[5]   M.A. Susner, Y. Yang, M.D. Sumption, E.W. Collings, M.A. Rindfleisch, M.J. Tomsic, and J.V. Marzik, *Supercond. Sci. Technol.* **24**(1), 012001/1-5 (2011).

[6]   W Goldacker, S.I. Schlachter, S. Zimmer, and H. Reiner, *Supercond. Sci. Technol.* **14** 787-793 (2001).

[7]   K. Tanaka, M. Okada, H. Kumakura, H. Kitaguchi, and K. Togano, *Physica C* **382**(2-3), 203-203 (2002).

[8]   H. Yamada, M. Igarashi, H. Kitaguchi, A. Matsumoto, and H. Kumakura, *Supercond. Sci. Technol.* **22** 075005/1-5 (2009).

[9]   X. Xu, J.H. Kim, S.X. Dou, S. Choi, J.H. Lee, H.W. Park, M.A. Rindfleisch, and M.J. Tomsic, *J. Appl. Phys.* 105(10), 103913/1-5 (2009).

[10]  S. Soltanian, X.L. Wang, J. Horvat, S.X. Dou, M.D. Sumption, M. Bhatia, E.W. Collings, P. Munroe, and M.J. Tomsic, *Supercond. Sci. Technol.* **18** 658-666 (2005).

[11**]**  V. Braccini, D. Nardelli, R. Penco, and G. Grasso, *Physica C* **456**(1-2), 209-217 (2007).

[12]  A. Kario, R. Nast, W. Häßler, C. Rodig, C. Mickel, W. Goldacker, B. Holzapfel, and L. Schultz, *Supercond. Sci. Technol.* **24** 075011/1-7 (2011).

[13]   M. Bhatia, M.D. Sumption, S. Bohnenstiehl, S.A. Dregia, E.W. Collings, M.J. Tomsic, and M.A. Rindfleisch, *IEEE Trans. Appl. Supercond.* **17**(2), 2750-2753 (2007).

[14]  C.E.J. Dancer, P. Mikheenko, A. Bevan, J.S. Abell, R.I. Todd, and C.R.M. Grovenor, *J. Eur. Cer. Soc.* **29**(9), 1817-1824 (2009).

[15] Y. Zhao, M. Ionescu, J. Horvat, A.H. Li, and S.X. Dou, *Supercond. Sci Technol*. **17**, 1247-1252 (2007).

[16] S.K. Chen, B.A. Glowacki, J.L. MacManus-Driscoll, M.E. Vickers, and M. Majoros, *Supercond. Sci Tech*. **17**, 243-248 (2004).

[17] X.F. Rui, Y. Zhao, Y.Y. Xu, L. Zhang, X.F. Sun, Y.Z. Wang, and H. Zhang, *Supercond. Sci Tech*. **17**, 689-691 (2004).

[18] S.K. Chen, M. Wei, and J.L. MacManus-Driscoll, *Appl. Phys. Lett*. **88**, 192512/1-3 (2006).

[19] C.H. Cheng, H. Zhang, Y. Zhao, Y. Feng, X.F. Rui, P. Munroe, H.M. Zeng, N. koshizuka, and M. Murakami, *Supercond. Sci. Technol.* **16**, 1182-1186 (2003).

[20] S. Okayasu, M. Sasase, K. Hojou, Y. Chimi, A. Iwase, H. Ikeda, R. Yoshizaki, T. Kambara, H. Sato, Y. Hamatani, and A. Maeda, *Physica C* **382** 104–107 (2002).

[21] M. Bhatia, M.D. Sumption, E.W. Collings, and S. Dregia, *Appl. Phys. Lett*. **87**(4), 042505/1-042505/3 (2005).

[22] X. Zhang, Z. Gao, D. Wang, Z. Yu, Y. Ma, S. Awaji and K. Watanabe, *Appl. Phys. Lett*. **89**, 132510/1-3 (2006).

[23] A. Agostino, M. Panetta, P. Volpe, M. Truccato, S. Cagliero, L. Gozzelino, R. Gerbaldo, G. Ghigo, F. Laviano, G. Lopardo, and B. Minetti, , *IEEE Trans. Appl. Supercond*. **17**(2), 2774-2777 (2007).

[24] J.S. Slusky, N. Rogado, K.A. Regan, M.A. Hayward, P. Khalifah, T. He, K. Inumaru, S.M. Loureiro, M.K. Haas, H.W. Zandbergen, and R.J. Cava, *Nature* **410**, 343-345 (2001).

[25] Y. Sun, D. Yu, Z. Liu, J. He, X. Zhang, Y. Tian, J. Xiang, and D. Zheng, *Appl. Phys. Lett*. **90**, 052507/1-3 (2007).

[26] N. Suemitsu, T. Masui, S. Lee, and S. Tajima, *Physica C* **445-448**, 39-41 (2006).

[27] K.M. Elsabawy and E.E. Kandyel, *Mater. Res. Bullet.* **42**, 1051-1060 (2007).

[28] W.K Yeoh and S.X. Dou, *Physica C* **456**, 170-179 (2007).

[29] Z.X. Shi, M.A. Susner, M.D. Sumption, E.W. Collings, X. Peng, M.A. Rindfleisch, and M.J. Tomsic, *Supercond. Sci. Technol*. **24,** 065015/1-7 (2011).

[30] M.A. Susner, M.D. Sumption, M. Bhatia, X. Peng, M.J. Tomsic, M.A. Rindfleisch, and E.W. Collings, *Physica C* **456,** 180–187 (2007).

[31] S.D. Bohnenstiehl, M.A. Susner, Y. Yang, E.W. Collings, M.D. Sumption, M.A. Rindfleisch , and R. Boone, *Physica C* **471** 108–111 (2011).

[32] R. H. T. Wilke, S. L. Bud'ko, P. C. Canfield D. K. Finnemore, Raymond J. Suplinskas, and S. T. Hannahs, *Phys. Rev. Lett.* **92**(21), 217003/1-4 (2004).

[33] A. Serquis, Y.T. Zhu, E.J. Peterson, J.Y. Coulter, D.E. Peterson, and F.M. Mueller, *Appl. Phys. Lett.* **79**(26)**,** 4399–4401 (2001).

[34] W.K. Yeoh, R.K. Zheng, S.P. Ringer, W.X. Li, X. Xu, S.X. Dou, S.K. Chen, and J.L. MacManus-Driscoll, *Scripta Materialia* **64**(4), 323-326 (2011).

[35] M. A. Susner, M. Bhatia, M. D. Sumption, and E. W. Collings, *J. Appl. Phys.* **105**, 103916/1-7 (2009).

[36] M. A. Susner, M. D. Sumption, M. Bhatia, M. J. Tomsic, M.A. Rindfleisch, and E. W. Collings, *Adv. Cryo. Eng*. **54,** 375-381 (2008).

[37] T. Matsushita, M. Kiuchi, A. Yamamoto, J. Shimoyama, and K. Kishio, *Supercond Sci. Technol.* **21**, 015008/1-7 (2008).

[38] M.S.A. Hossain, C. Senatore, R. Flükiger, M.A. Rindfleisch, M.J. Tomsic, J.H. Kim, and S.X. Dou, *Supercond. Sci. Technol.* **22,** 095004/1-8 (2009).

[39] J.M. Rowell, *Supercond. Sci. Technol*. **16,** R17–R27 (2003).

[40] T. Matsushita, M. Kiuchi, A. Yamamoto, J. Shimoyama, and K. Kishio, Physica C **468**, 1833-1835 (2008).

[41] R.F. Klie, J.C. Idrobo, N.D. Browning, K.A. Regan, N.S. Rogado, and R.J. Cava, *Appl. Phys. Lett*. **79**(12), 1837-1839 (2001).

[42] B. Birajdar, N. Peranio, and O. Eibl, *Supercond. Sci. Tech* **21**, 1-20 (2008).

[43] D. Uchiyama, K. Mizuno, T. Akao, M. Maeda, T. Kawakami, H. Kobayashi, Y. Kubota, and K. Yasohama *Cryogenics* **47**, 282-286 (2007).

[44] E. Bartolomé, F. Gömöry, X. Granados, T. Puig, and X. Obradors, *Supercond. Sci. Techol.* **18**, 388-394 (2005).

[45] J. Horvat, W.K. Yeoh, J.H. Kim, and S.X. Dou. *Supercond. Sci. Tech.* **21** 065003/1-6 (2008).

[46] M.D. Sumption, *Applied Superconductivity* **2**(1) 41-46 (1994).

[47]   Y. Zhang, C. Lu, S. Zhou, and J. Joo, *J. Nanosci. Nanotechnol.* **9** 7402-7406 (2009).

[48]   M.D. Sumption, M.A. Susner, M. Bhatia, M.A. Rindfleisch, M.J. Tomsic, K.J. McFadden, and E.W. Collings *IEEE Trans. Appl. Supercond.* **17,** 2838–41 (2007).

[49]   M.J. Tomsic, M.A. Rindfleisch, J. Yue, K.J. McFadden, D. Doll, J. Phillips, M.D. Sumption, S. Bohnenstiehl, and E.W. Collings *Physica* C **456** 203–8 (2007).

[50]   M.D. Sumption, M. Bhatia, M.A. Rindfleisch, M.J. Tomsic, and E.W. Collings *Appl. Phys. Lett.* **86** 092507/1-3 (2005).

[51]   G. Giunchi, S. Ceresara, G. Ripamonti, A. Di Zenobio, S. Rossi, S. Chiarelli, M. Spadoni, R. Wesche, and P.L. Bruzzone *Supercond. Sci. Technol.* **16** 285-291 (2003).

[52]   G. Giunchi, G. Ripamonti, E. Perini, T. Cavallin, and E. Bassani *IEEE Trans. Appl. Supercond.* **17**(2) 2761–2765 (2007).

[53]   J.M. Hur, K. Togano, A. Matsumoto, H. Kumakura, H. Wada, and K. Kimura *Supercond. Sci.Technol.* **21** 032001/1-4 (2008).

[54]   J.D. DeFouw and D.C. Dunand *Acta Materialia* **56,** 5751–5763 (2008)

[55]   Z.X. Shi, M. Tokunaga, A.K. Pradham, T. Tamegai, Y. Takano, K. Togano, H. Kito, and H. Ihara, *Physica C* **370** 6-12 (2002).

[56]   D. Pal, L. DeBeer-Schmitt, T. Bera, R. Cubitt, C.D. Dewhurst, J. Jun, N.D. Zhigadlo, J. Karpinski, V.G. Kogan, and M.R. Eskildsen, *Phys. Rev. B* **73** 012513/1-4 (2006).

[57]   J.G. Byrne, *Recovery, Recrystallization, and Grain Growth*, New York: The MacMillan Co., 1965, p.99.

[58]   X. Song, V. Braccini, and D.C. Larbalestier, *J. Mater. Res.* **19** 2245-2255 (2004)

[59]   J. Jiang, B.J. Senkowitz, D.C. Larbalestier, and E.E. Hellstrom, *Supercond. Sci. Technol.* **19** L33-L36 (2006)

[60]   E.J. Kramer, *J. Appl. Phys*. **44**(3) 1360-1370 (1973).

[61]   D. Dew-Hughes*, Phil. Mag*. **30**, 293-305 (1974).

[62]   T. Matsushita, *Flux Pinning in Superconductors,* New York: Springer, 2007, p. 416.

[63]   E. H. Brandt, *Phys. Rev. B* **58** 6506-6522 (1998).

[64]   M.J. Qin, S. Keshavarzi, S. Soltanian, X.L. Wang, H.K. Liu, and S.X. Dou, *Phys.Rev. B* **69** 012507/1-4 (2004).

[65]   S. Soltanian, M.J. Qin, S. Keshavarsi, X.L. Wang, and S.X. Dou, *Phys. Rev. B* **68** 134509/1-4 (2003).

[66]   J. Horvat, S. Soltanian, X.L. Wang, and S.X. Dou, *Applied Physics Letters*, **84** 3109- 3111(2004).

[67]   E.W. Collings, M.D. Sumption, M. Bhatia, M.A. Susner, and S.D. Bohnenstiehl, *Supercond. Sci. Tech*. **21**(10), 103001/1-4 (2008).

**List of Tables**

Table I. Definitions of Symbols for Critical Current Density, CCD

| Symbol | Definition |
|---|---|
| $J_{ct}$ | Measured transport CCD |
| $J_{cm}^{\perp}$ | Measured perpendicular field magnetic CCD |
| $J_{cm}^{\phi}$ | Measured parallel field magnetic CCD |
| $J_{cm,\ MOD}^{\perp}$ | Model perpendicular field magnetic CCD, Eqn. (1) |
| $J_3$ | Model strand-longitudinal CCD, Eqn. (1), $= J_{ct}$ |
| $J_2$ | Model strand-transverse CCD, Eqn. (1), $= J_{cm}^{\phi}$ |

**List of Figures**

Figure 1.  Conceptualization of Mg elongation during the wire drawing process.

Figure 2.  BSE SEM images of the longitudinal cross sections of an unreacted $MgB_2$ strand at various stages along the wire reduction path.  Evident is the evolution of macrostrure, in particular the ribbonization of the Mg. The strand (starting OD of 9.53 mm) is shown at wire OD/strain values of (a) 4.08 mm /57% area reduction, (b) 2.41 mm/75% area reduction, (c) 1.42 mm/85% area reduction, and (d) 0.83 mm/91% area reduction.

Figure 3.  Aspect ratio of Mg stringers vs. % area reduction of the $MgB_2$ strand.

Figure 4.  BSE performed on longitudinal cross sections of strand HT for various times at $600^{o}C$ and $700^{o}C$ showing macrostructural evolution of phases with HT.

Figure 5.  $\chi/\chi_0$ vs. $T$ and $d(\chi/\chi_0)/dT$ vs. $T$ for four HT times at $600^{o}C$ and $700^{o}C$.

Figure 6.  Onset $T_c$ vs. HT time at of $600^{o}C$ and $700^{o}C$.

Figure 7.  4.2 K transport $J_c$ for HT temperatures of $600^{o}C$ and $700^{o}C$.

Figure 8.  5 K, 4 T values of $J_{cm}^{\perp}$ as functions of HT time at $600^{o}C$ and $700^{o}C$.  Lines are plotted as guides to the eye.

Figure 9.  Fracture UHR-SEMs showing a gradual growth of grains from 35~55 nm with HT time.

Figure 10.  Grain size vs. HT time for $600^{o}C$ and $700^{o}C$.  The solid and dashed lines represent fittings to the standard grain growth relationship $d \propto t^{1/2}$.

Figure 11.  a) Grain size ($d$) and 5 K, 4 T $J_{cm}^{\perp}$ vs. HT time for the $700^{o}C$ HT; and b) 5 K, 4 T $F_p$ vs. inverse grain size.

Figure 12.  Measured $J_{ct}$ together with measured $J_{cm}^{\perp}$ at aspect ratios of $S=8$ and $S=13$.

Figure 13.  $J_{cm}^{\perp}{}_{MOD}$ (Equations 1a and 1b) as a function of $B$ compared to $J_{ct}$, $J_{cm}^{\phi}$ and $J_{cm}^{\perp}$.  The inset shows the Kramer plot of the same $J_c$-$B$ curves to emphasize the different values of $B_{irr}$ that can be measured depending on the technique.

Figure 14.  a) Unreacted Mg+B wire, longitudinal cross section; b) long dashed lines emphasizing the continuous nature of the longitudinal boron veins; and c) short dashed lines showing the discontinuous and circuitous nature of the transversal boron connections.

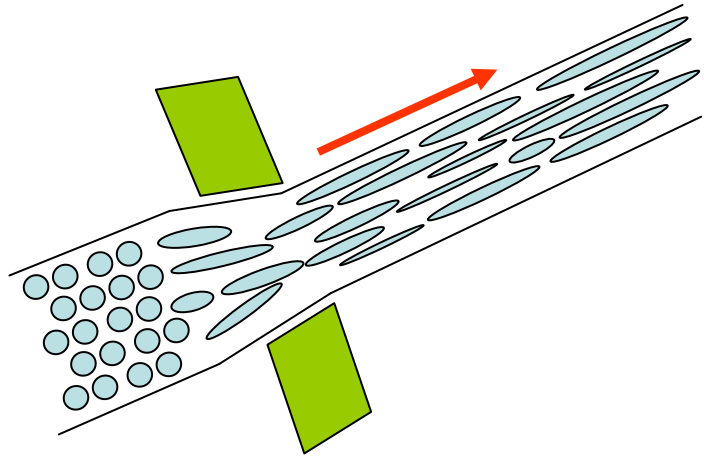

Figure 1. Conceptualization of Mg elongation during the wire drawing process.

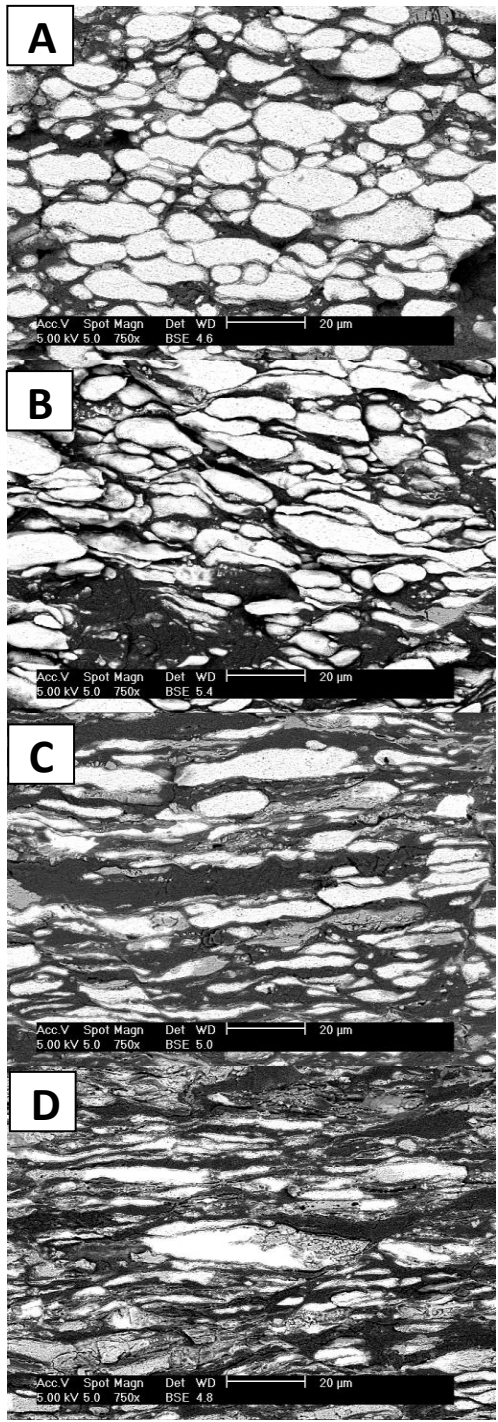

Figure 2. BSE SEM images of the longitudinal cross sections of an unreacted $MgB_2$ strand at various stages along the wire reduction path. Evident is the evolution of macrostrure, in particular the ribbonization of the Mg. The strand (starting OD of 9.53 mm) is shown at wire OD/strain values of (a) 4.08 mm /57% area reduction, (b) 2.41 mm/75% area reduction, (c) 1.42 mm/85% area reduction, and (d) 0.83 mm/91% area reduction.

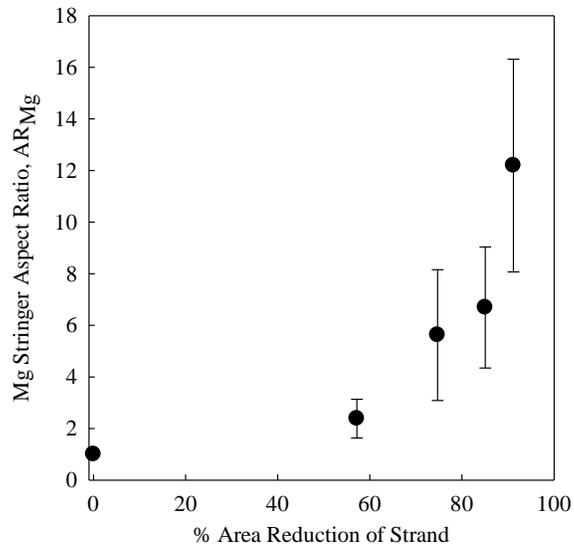

Figure 3. Aspect ratio of Mg stringers vs. % area reduction of the $MgB_2$ strand.

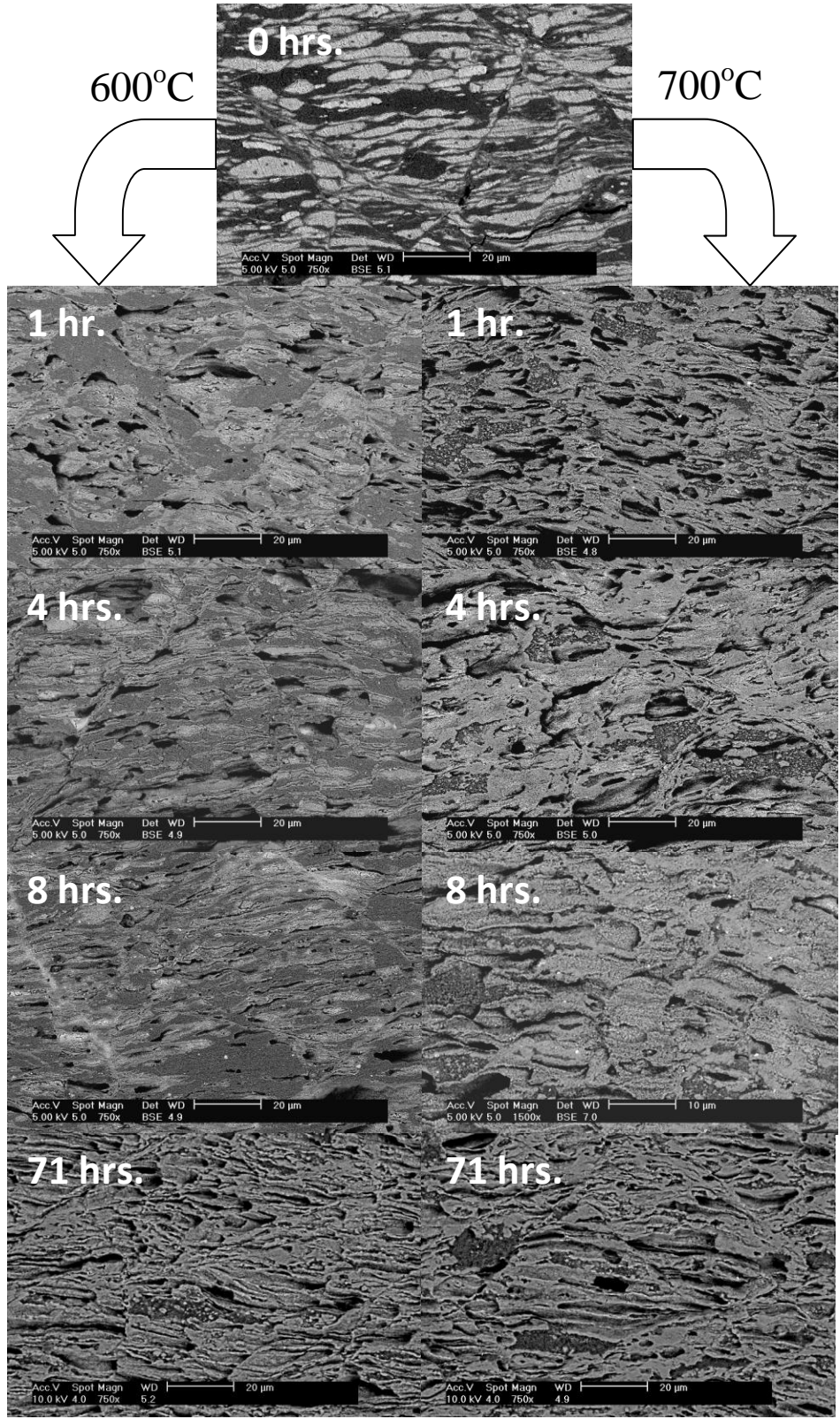

Figure 4. BSE performed on longitudinal cross sections of strand HT for various times at 600°C and 700°C showing macrostructural evolution of phases with HT.

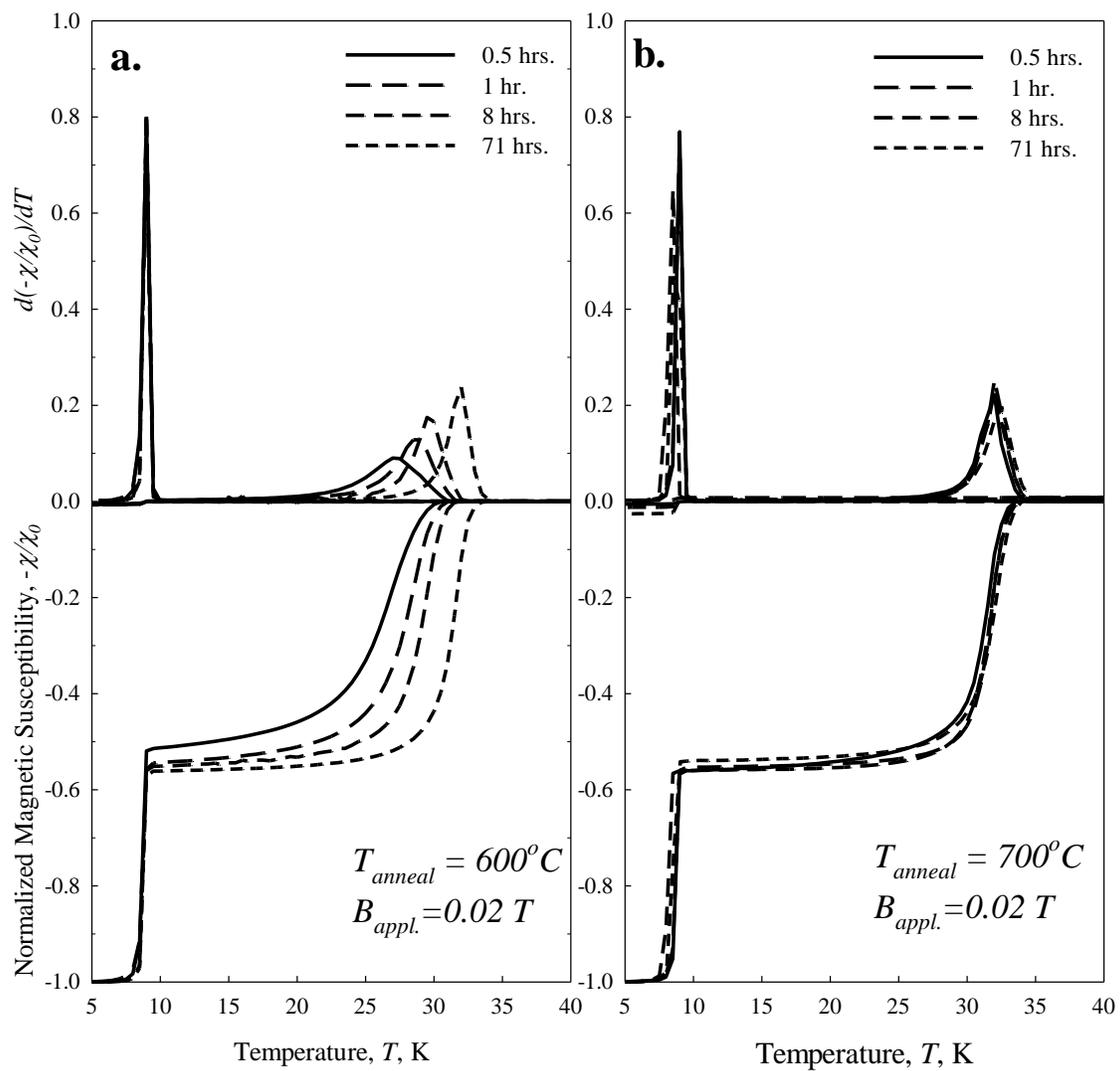

Figure 5. $\chi/\chi_0$ vs. $T$ and $d(\chi/\chi_0)/dT$ vs. $T$ for four HT times at 600°C and 700°C.

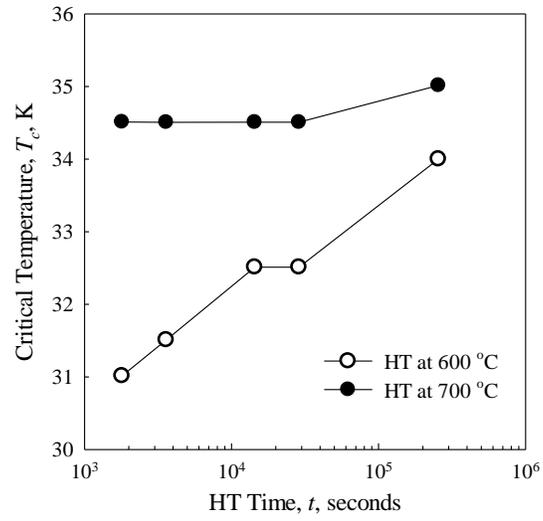

Figure 6. Onset $T_c$ vs. HT time at of 600°C and 700°C.

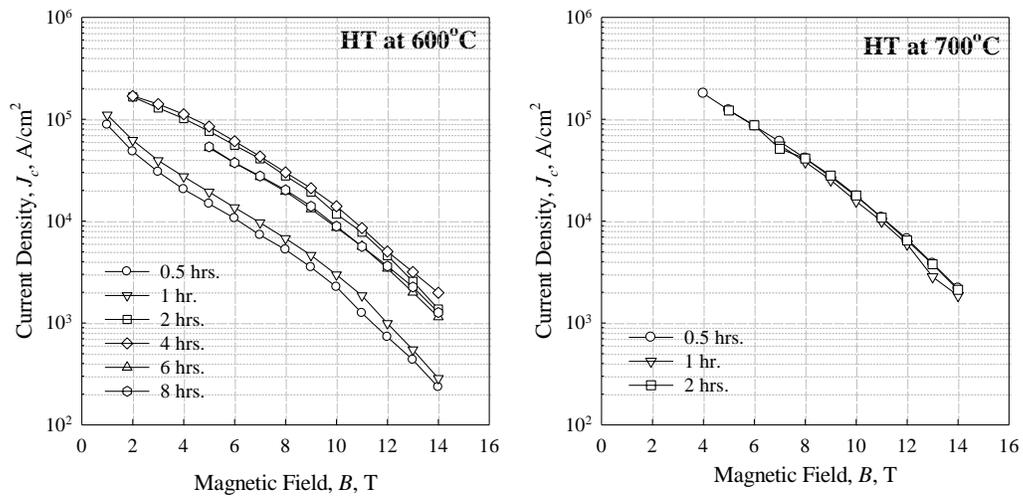

Figure 7.  4.2 K transport $J_c$ for HT temperatures of 600°C and 700°C.

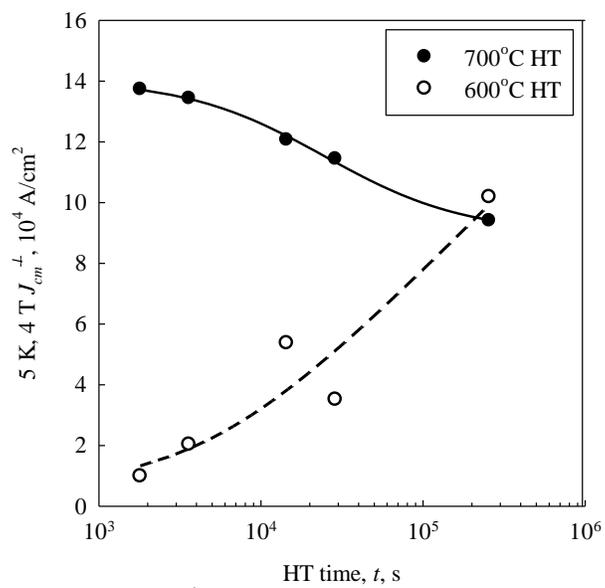

Figure 8. 5 K, 4 T values of $J_{cm}^{\perp}$ as functions of HT time at 600°C and 700°C. Lines are plotted as guides to the eye.

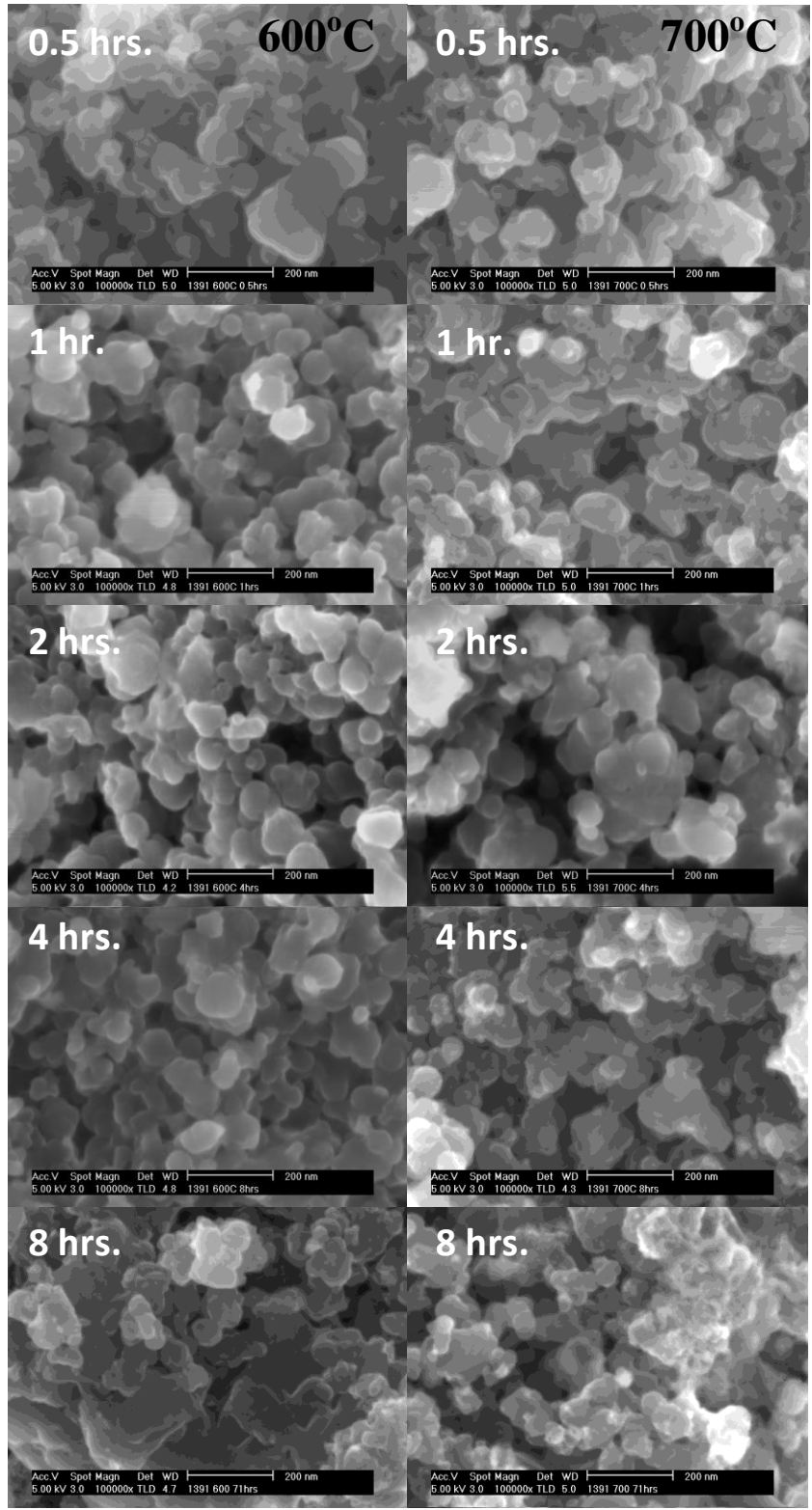

Figure 9. Fracture UHR-SEMs showing a gradual growth of grains from 35~55 nm with HT time.

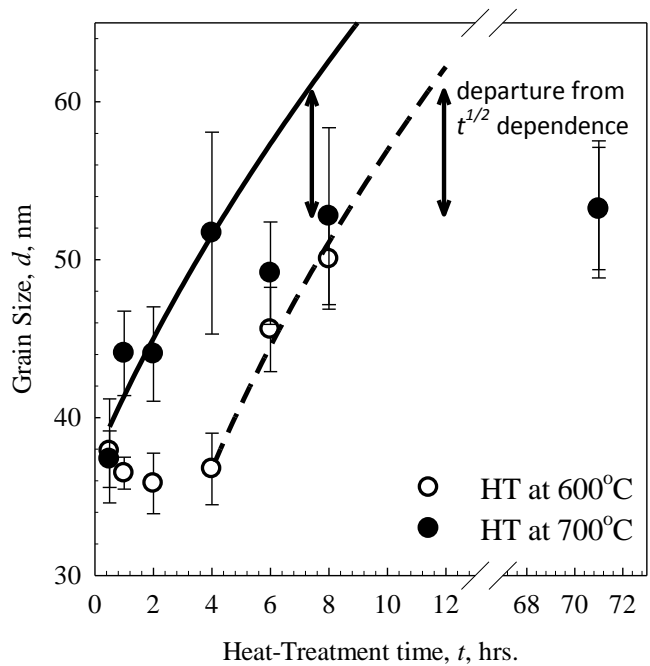

Figure 10. Grain size vs. HT time for 600°C and 700°C. The solid and dashed lines represent fittings to the standard grain growth relationship $d \propto t^{1/2}$.

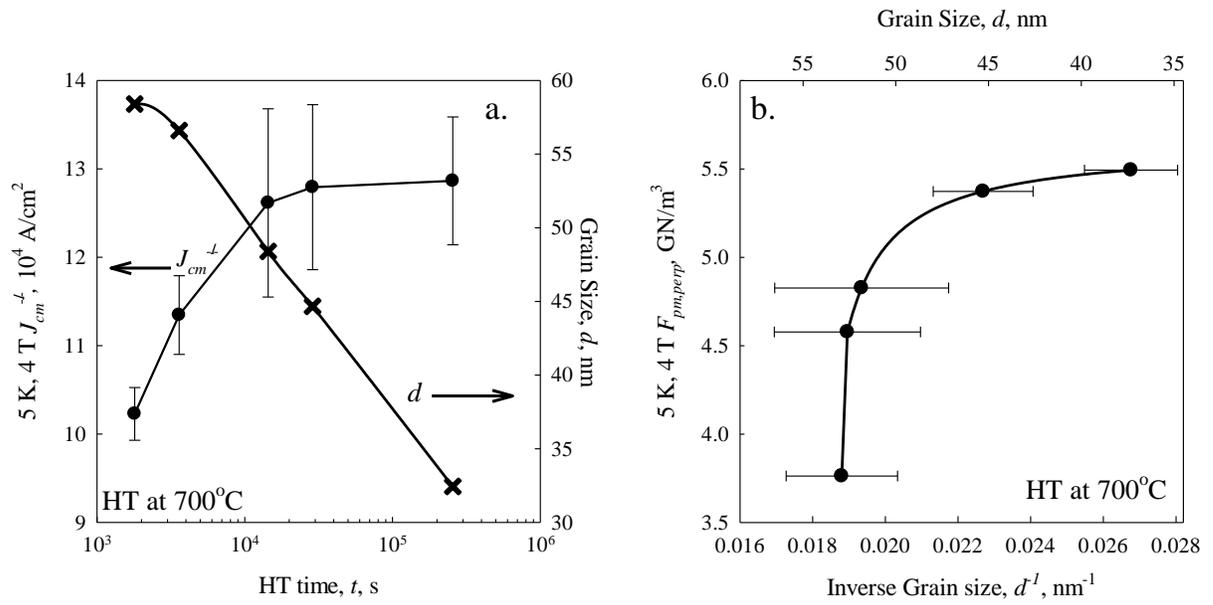

Figure 11. a) Grain size ($d$) and 5 K, 4 T $J_{cm}^{\perp}$ vs. HT time for the 700°C HT; and b) 5 K, 4 T $F_p$ vs. inverse grain size.

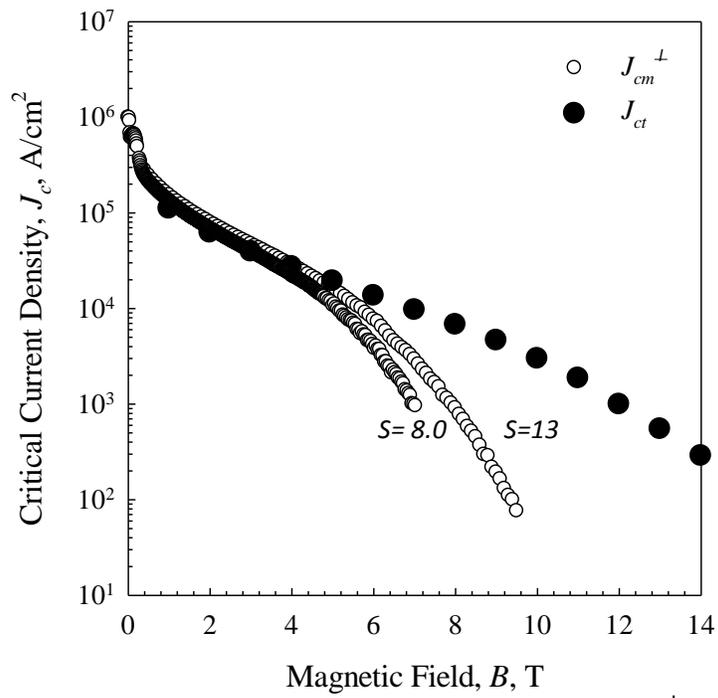

Figure 12. Measured $J_{ct}$ together with measured $J_{cm}^{\perp}$ at aspect ratios of $S=8$ and $S=13$.

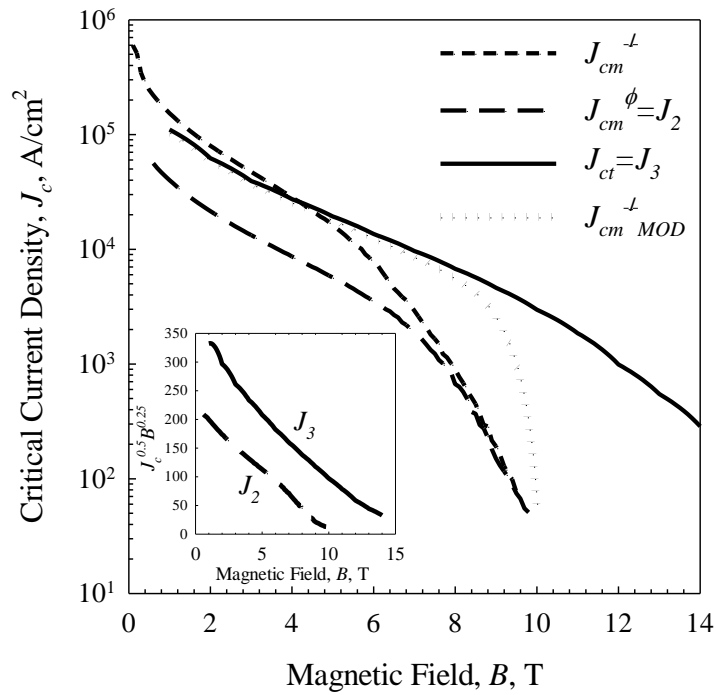

Figure 13. $J_{cm}^{\perp}{}_{MOD}$ (Equations 1a and 1b) as a function of $B$ compared to $J_{ct}$, $J_{cm}^{\phi}$ and $J_{cm}^{\perp}$. The inset shows the Kramer plot of the same $J_c$-$B$ curves to emphasize the different values of $B_{irr}$ that can be measured depending on the technique.

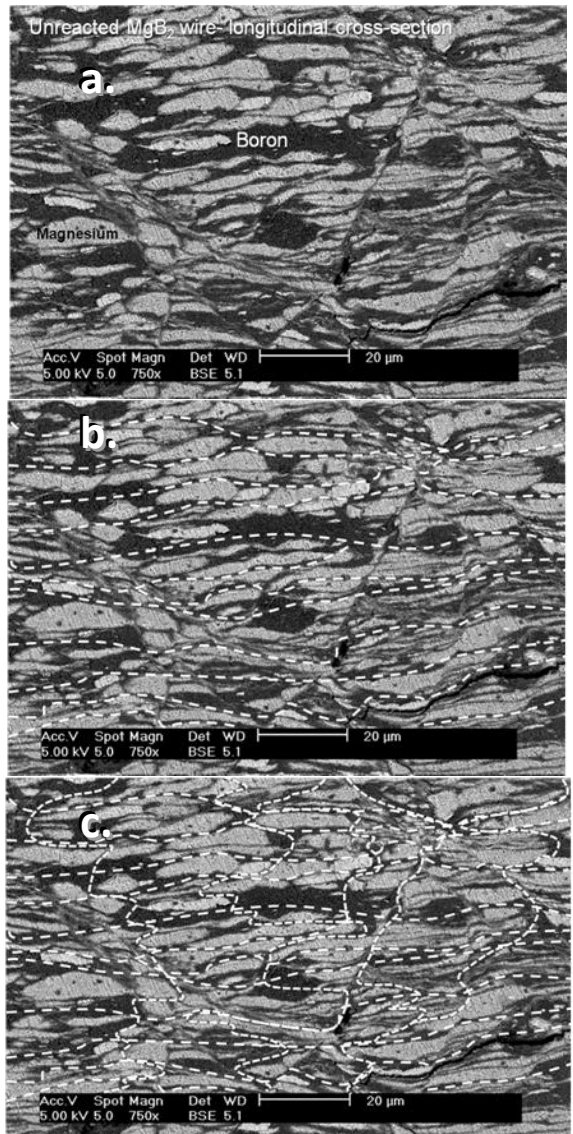

Figure 14. a) Unreacted Mg+B wire, longitudinal cross section; b) long dashed lines emphasizing the continuous nature of the longitudinal boron veins; and c) short dashed lines showing the discontinuous and circuitous nature of the transversal boron connections.